\documentclass[epjc3,twocolumn,a4paper,fleqn]{svjour3} 
\usepackage[fleqn]{amsmath}
\usepackage{graphicx}
\usepackage{epstopdf}
\usepackage{fix-cm}
\usepackage{txfonts}
\usepackage{lipsum}
\usepackage{mathtools}
\usepackage{cuted}
\usepackage{multirow}
\usepackage{comment}
\usepackage{empheq}
\usepackage[titletoc,title]{appendix}

\usepackage{listings}
\usepackage{lineno}
\usepackage{mdwlist}
\usepackage{pennames}
\usepackage[british]{babel}
\usepackage{color}
\usepackage{subfigure}
\usepackage{siunitx}
\usepackage{import}
\usepackage{amssymb}
\usepackage{epsfig}
\usepackage{cite}
\usepackage{hyperref}
\hypersetup{
colorlinks=true,
linkcolor=blue,
citecolor=blue,
urlcolor=blue,
}

\DeclareSIUnit \clight  {\textit{c}}
\RequirePackage[normalem]{ulem} 
\RequirePackage{color}\definecolor{RED}{rgb}{1,0,0}\definecolor{BLUE}{rgb}{0,0,1} 
\RequirePackage[normalem]{ulem} 
\RequirePackage{color}\definecolor{RED}{rgb}{1,0,0}\definecolor{BLUE}{rgb}{0,0,1} 

\DeclareGraphicsExtensions{.eps,.pdf,.jpeg,.jpg,.png}

\def\vector#1{\mbox{\boldmath $#1$}}

\newcommand{\bea}{\begin{eqnarray}}
\newcommand{\eea}{\end{eqnarray}}
\newcommand{\be}{\begin{equation}}
\newcommand{\ee}{\end{equation}}
\newcommand{\fref}[1]{Fig.~\ref{#1}}

\newcommand{\sref}[1]{Sect.~\ref{#1}}

\newcommand{\tref}[1]{Table~\ref{#1}}

\newcommand*{\muonp}          {\ifmmode\mathrm{\muup^+}\else$\mathrm{\muup^+}$\fi}
\newcommand*{\muon}           {\ifmmode\mathrm{\muup}\else$\mathrm{\muup}$\fi}
\newcommand*{\tauon}          {\ifmmode\mathrm{\tauup}\else$\mathrm{\tauup}$\fi}

\newcommand*{\egamma}         {E_{\mathrm{\gammaup}}}
\newcommand*{\photon}         {\ifmmode{\gammaup}\else${\gammaup}$\fi}
\newcommand*{\positron}       {\ifmmode{\mathrm{e}^+}\else${\mathrm{e}^+}$\fi}
\newcommand*{\electron}       {\ifmmode{\mathrm{e}}\else${\mathrm{e}}$\fi}
\newcommand*{\epositron}      {{E_\mathrm{e^+}}}

\newcommand*{\tpositron}      {{t_\mathrm{e^+}}}

\newcommand*{\tgamma}         {{t_{\mathrm{\gammaup}}}}
\newcommand*{\tegamma}        {{t_{\mathrm{e^+ \gammaup}}}}
\newcommand{\tg}{{\ifmmode t_{\gammaup_1\mathrm{e}^+}\else$t_{\gammaup_1\mathrm{e}^+}$\fi}}
\newcommand*{\tgg}{{\ifmmode t_{\gammaup\gammaup}\else$t_{\gammaup\gammaup}$\fi}}
\newcommand*{\Thetaegamma}    {{\Theta_{\mathrm{e}^+ \gammaup}}}

\newcommand*{\thetaegamma}    {{\theta_{\mathrm{e}^+ \gammaup}}}
\newcommand*{\phiegamma}      {{\phi_{\mathrm{e}^+ \gammaup}}}
\newcommand*{\thetae}         {{\theta_\mathrm{e^+}}}
\newcommand*{\phie}           {{\phi_\mathrm{e^+}}}
\newcommand*{\thetagamma}     {{\theta_\mathrm{\gammaup}}}
\newcommand*{\phigamma}       {{\phi_\mathrm{\gammaup}}}

\newcommand{\meg}{\ifmmode{\muup \to e \gammaup}\else$\mathrm{\muup \to e \gammaup}$\fi}
\newcommand{\megp}{\ifmmode{\muup^+ \to \mathrm{e}^+ \gammaup}\else$\mathrm{\muup^+ \to e^+ \gammaup}$\fi}
\newcommand{\michel}{\ifmmode{\muup^+ \to e^+ \nuup\bar{\nuup}}\else$\mathrm{\muup^+ \to e^+ \nuup\bar{\nuup}}$\fi}
\newcommand{\radiative}{\ifmmode{\muup^+ \to \mathrm{e}^+\nuup\bar{\nuup}\gammaup} \else$\mathrm{\muup^+ \to e^+ \nuup\bar{\nuup}\gammaup}$\fi}
\newcommand{\conv}{\ifmmode{\muup^- \to e^-}\else$\mathrm{\muup^- \to e^-}$\fi}
\newcommand{\convN}{\ifmmode{\muup^-N \to e^-N}\else$\mathrm{\muup^-N \to e^-N}$\fi}
\newcommand{\mute}{\ifmmode{\muup \to 3e}\else $\mathrm{\muup \to 3e}$\fi}
\newcommand{\mutec}{\ifmmode{\muup^+ \to e^+e^+e^-}\else $\mathrm{\muup^+ \to e^+e^+e^-}$\fi}
\newcommand{\aif}{\ifmmode\mathrm{e}^+ \mathrm{e}^- \to \gammaup\gammaup \else$\mathrm{e}^+ \mathrm{e}^- \to \gammaup \gammaup$\fi}
\newcommand{\teg}{\ifmmode{\tauup \to e \gammaup} \else$\mathrm{\tauup \to e \gammaup}$\fi}
\newcommand{\tmg}{\ifmmode{\tauup \to \gammaup} \else$\mathrm{\tauup \to \muup \gammaup}$\fi}
\newcommand{\tmueg}{\ifmmode{\mathrm\tauup \to \ell \gammaup}\else$\mathrm{\tauup \to \ell \gammaup}$\fi}
\newcommand{\tautl}{\ifmmode{\mathrm\tauup \to 3\ell} \else$\mathrm\tauup \to 3\ell$\fi}

\newcommand*{\BR}     { {\cal B} }

\newcommand*{\ypos}          {y_\mathrm{e^+}}
\newcommand*{\zpos}          {z_\mathrm{e^+}}

\newcommand*{\ugamma}         {u_{\gammaup}}
\newcommand*{\vgamma}         {v_{\gammaup}}
\newcommand*{\wgamma}         {w_{\gammaup}}

\newcommand*{\nsig}           {N_{\rm sig}}
\newcommand*{\nrd}            {N_{\rm RMD}}
\newcommand*{\nbg}            {N_{\rm BG}}
\newcommand*{\nacc}            {N_{\rm ACC}}
\newcommand*{\xt}            {x_{\rm T}}

\newcommand*{\sens}     { {\cal S}_{90}}
\newcommand*{\ul}     { {\cal B}_{90}}
\newcommand*{\bestfit}     { {\cal B}_\mathrm{fit}}
\newcommand*{\rsig}           {R_{\rm sig}} 

\newcommand*{\mathtentative}{}
\def\mathtentative#1#{\mathcoloraux{#1}}
\newcommand*{\mathcoloraux}[3]{%
  \protect\leavevmode
  \begingroup
    \color#1{#2}#3%
  \endgroup
}

\journalname{Eur. Phys. J. C} 

\begin{document}



\title{New limit on the \megp\ decay with the MEG II experiment
}
\author{The MEG~II collaboration}
\newcommand*{\INFNPi}{INFN Sezione di Pisa$^{a}$; Dipartimento di Fisica$^{b}$ dell'Universit\`a, Largo B.~Pontecorvo~3, 56127 Pisa, Italy}
\newcommand*{\INFNGe}{INFN Sezione di Genova$^{a}$; Dipartimento di Fisica$^{b}$ dell'Universit\`a, Via Dodecaneso 33, 16146 Genova, Italy}
\newcommand*{\INFNPv}{INFN Sezione di Pavia$^{a}$; Dipartimento di Fisica$^{b}$ dell'Universit\`a, Via Bassi 6, 27100 Pavia, Italy}
\newcommand*{\INFNRm}{INFN Sezione di Roma$^{a}$; Dipartimento di Fisica$^{b}$ dell'Universit\`a ``Sapienza'', Piazzale A.~Moro, 00185 Roma, Italy}
\newcommand*{\INFNNa}{INFN Sezione di Napoli, Via Cintia, 80126 Napoli, Italy}
\newcommand*{\INFNLe}{INFN Sezione di Lecce$^{a}$; Dipartimento di Matematica e Fisica$^{b}$ dell'Universit\`a, Via per Arnesano, 73100 Lecce, Italy}
\newcommand*{\ICEPP} {ICEPP, The University of Tokyo, 7-3-1 Hongo, Bunkyo-ku, Tokyo 113-0033, Japan }
\newcommand*{\Kobe} {Kobe University, 1-1 Rokkodai-cho, Nada-ku, Kobe, Hyogo 657-8501, Japan}
\newcommand*{\UCI}   {University of California, Irvine, CA 92697, USA}
\newcommand*{\KEK}   {KEK, High Energy Accelerator Research Organization, 1-1 Oho, Tsukuba, Ibaraki 305-0801, Japan}
\newcommand*{\PSI}   {PSI Center for Neutron and Muon Sciences, 5232 Villigen, Switzerland}
\newcommand*{\Waseda}{Research Institute for Science and Engineering, Waseda~University, 3-4-1 Okubo, Shinjuku-ku, Tokyo 169-8555, Japan}
\newcommand*{\BINP}  {Budker Institute of Nuclear Physics of Siberian Branch of Russian Academy of Sciences, 630090 Novosibirsk, Russia}
\newcommand*{\JINR}  {Joint Institute for Nuclear Research, 141980 Dubna, Russia}
\newcommand*{\ETHZ}  {Institute for Particle Physics and Astrophysics, ETH Z\" urich, 
Otto-Stern-Weg 5, 8093 Z\" urich, Switzerland}
\newcommand*{\NOVST} {Novosibirsk State Technical University, 630092 Novosibirsk, Russia}
\newcommand*{\ScuolaPi}{Scuola Normale Superiore, Piazza dei Cavalieri 7, 56126 Pisa, Italy}
\newcommand*{\INFNLNF}{\textit{Present Address}: INFN, Laboratori Nazionali di Frascati, Via 
E. Fermi, 40-00044 Frascati, Rome, Italy}
\newcommand*{\Liverpool}{Oliver Lodge Laboratory, University of Liverpool, Liverpool, L69 7ZE, United Kingdom}

\date{Received: date / Accepted: date}

\author{
The MEG~II collaboration\\\\
        K.~Afanaciev~\thanksref{addr12} \and
        A.~M.~Baldini\thanksref{addr1}$^{a}$ \and
        S.~Ban~\thanksref{addr10} \and
        H.~Benmansour\thanksref{addr1}$^{ab}$ \and
        G.~Boca~\thanksref{addr4}$^{ab}$ \and        P.~W.~Cattaneo~\thanksref{addr4}$^{a}$\thanksref{e1} \and
        G.~Cavoto~\thanksref{addr5}$^{ab}$ \and
        F.~Cei~\thanksref{addr1}$^{ab}$ \and
        M.~Chiappini~\thanksref{addr1}$^{ab}$ \and
        A.~Corvaglia~\thanksref{addr6}$^{a}$ \and
        G.~Dal~Maso\thanksref{addr2,addr16} \and
        A.~De~Bari~\thanksref{addr4}$^{a}$ \and
        M.~De~Gerone~\thanksref{addr3}$^{a}$ \and
        L.~Ferrari~Barusso~\thanksref{addr3}$^{ab}$ \and
        M.~Francesconi~\thanksref{addr17} \and 
        L.~Galli~\thanksref{addr1}$^{a}$ \and
        G.~Gallucci~\thanksref{addr3}$^{a}$ \and
        F.~Gatti~\thanksref{addr3}$^{ab}$ \and
        L.~Gerritzen~\thanksref{addr10}  \and
        F.~Grancagnolo~\thanksref{addr6}$^{a}$ \and
        E.~G.~Grandoni~\thanksref{addr1}$^{ab}$ \and 
        M.~Grassi~\thanksref{addr1}$^{a}$ \and 
        D.~N.~Grigoriev~\thanksref{addr7,addr8} \and
        M.~Hildebrandt~\thanksref{addr2} \and
        F.~Ignatov~\thanksref{addr15} \and
        F.~Ikeda~\thanksref{addr10}  \and
        T.~Iwamoto~\thanksref{addr10}  \and
        S.~Karpov~\thanksref{addr7} \and
        P.-R.~Kettle~\thanksref{addr2} \and
        N.~Khomutov~\thanksref{addr12} \and
        A.~Kolesnikov~\thanksref{addr12}  \and
        N.~Kravchuk~\thanksref{addr12}  \and
        V.~Krylov~\thanksref{addr12} \and
        N.~Kuchinskiy~\thanksref{addr12}  \and
        F.~Leonetti~\thanksref{addr1}$^{ab}$ \and
        W.~Li~\thanksref{addr10} \and
        V.~Malyshev~\thanksref{addr12}  \and
        A.~Matsushita~\thanksref{addr10}  \and
        S.~Mihara~\thanksref{addr13}  \and
        W.~Molzon~\thanksref{addr11} \and
        Toshinori~Mori~\thanksref{addr10}  \and
        D.~Nicol\`o~\thanksref{addr1}$^{ab}$ \and
        H.~Nishiguchi~\thanksref{addr13}  \and
        A.~Ochi~\thanksref{addr14}\thanksref{e5}  \and
        S.~Ogawa~\thanksref{addr10}  \and
        W.~Ootani~\thanksref{addr10}  \and
        A.~Oya~\thanksref{addr10} \and
        D.~Palo~\thanksref{addr11} \and
        M.~Panareo~\thanksref{addr6}$^{ab}$ \and
        A.~Papa~\thanksref{addr1}$^{ab}$\thanksref{addr2} \and
        D.~Pasciuto~\thanksref{addr5}$^{a}$ \and   
        A.~Popov~\thanksref{addr7} \and
        F.~Renga~\thanksref{addr5}$^{a}$ \and
        S.~Ritt~\thanksref{addr2} \and
        M.~Rossella~\thanksref{addr4}$^{a}$ \and
        A.~Rozhdestvensky~\thanksref{addr12}  \and
        S.~Scarpellini~\thanksref{addr5}$^{ab}$ \and
        G.~Signorelli~\thanksref{addr1}$^{a}$ \and
        H.~Suzuki~\thanksref{addr14}  \and
        M.~Takahashi~\thanksref{addr14}  \and
        Y.~Uchiyama~\thanksref{addr10,addr14} \and
        R.~Umakoshi~\thanksref{addr10} \and
        A.~Venturini~\thanksref{addr1}$^{ab}$ \and
        B.~Vitali~\thanksref{addr1}$^{a,}$\thanksref{addr5}$^{b}$ \and
        C.~Voena~\thanksref{addr5}$^{ab}$ \and   
        K.~Yamamoto~\thanksref{addr10}  \and
        R.~Yokota~\thanksref{addr10}  \and
        T.~Yonemoto~\thanksref{addr10}  
}

\institute{\JINR   \label{addr12}
           \and
             \INFNPi \label{addr1}
           \and
             \ICEPP \label{addr10}
            \and
             \INFNPv \label{addr4}
           \and
             \INFNRm \label{addr5}
           \and
             \INFNLe \label{addr6} 
           \and
             \PSI \label{addr2}
            \and
             \ETHZ \label{addr16}
           \and
             \INFNGe \label{addr3}
            \and
             \INFNNa \label{addr17} 
            \and
             \BINP   \label{addr7}
           \and
             \NOVST  \label{addr8}
         \and
             \Liverpool  \label{addr15}
           \and
             \UCI    \label{addr11}
           \and
             \KEK    \label{addr13}
           \and
             \Kobe    \label{addr14}
}

\thankstext[*]{e1}{Corresponding author: paolo.cattaneo@pv.infn.it} 
\thankstext[$\dagger $]{e5}{Deceased} 

\maketitle 

\begin{abstract}
This letter reports the result of the search 
for the decay \megp\ undertaken at the Paul 
Scherrer Institut in Switzerland with
the MEG~II experiment using the data collected 
in the 2021--2022 physics runs.
The sensitivity of this search is 
$\num{2.2e-13}$, a factor of 2.4 better than that 
of the full MEG dataset and obtained in a data taking period of about one fourth that of MEG, thanks to the superior performances of the new detector.
The result is consistent with the 
expected background, yielding an upper limit 
on the branching ratio of ${\cal B} (\megp) < \num{1.5e-13}$
(\SI{90}{\percent} C.L.).
Additional improvements are expected with the data 
collected during the years 2023--2024. 
The data-taking will continue in the coming years.
\end{abstract}


\tableofcontents 
In the standard model (SM) of particle physics, charged lepton flavour-violating (CLFV) processes are
almost forbidden, with extremely small branching ratio
(\num{\sim e-54} \cite{Petcov:1977} 
when considering non-zero neutrino mass differences 
and mixing angles. Hence, the experimental searches for such decays are free from SM contributions
and a positive signal would be unambiguous evidence for physics beyond the SM. Several SM extensions 
\cite{barbieri1994, Hisano:1997tc} predict experimentally accessible CLFV decay rates, with the channel \megp\ being particularly sensitive.

The most stringent upper limit on the \megp\ branching ratio was set by the MEG~II collaboration,
${\cal B} (\megp) < \SI{3.1e-13}{}$ (\SI{90}{\percent} C.L.) \cite{MEGSearch2023} 
by combining the full dataset of the MEG experiment 
\cite{baldini_2016} with the dataset of the MEG~II experiment collected in 2021. 
This letter presents a new limit on ${\cal B} (\megp)$ based on the analysis 
of the data collected in 2022 combined with an update analysis 
of the 2021 dataset.

\section{Signal and background}

A signal event is a back-to-back, monoenergetic, 
time coincident photon-positron pair from a decay \megp\ at rest. 
The energy of both photon and positron is half the muon mass (\SI{52.83}{\MeV}) for muons 
decaying at rest. 
The background consists of events from
radiative muon decay (RMD) \radiative\ 
and from accidental time coincidences of positrons
from muon Michel decay \michel\ with photons from RMD, 
positron-electron annihilation-in-flight or bremsstrahlung (ACC).

\section{The MEG~II experiment}

\begin{figure*}[tbp]
\centering
\includegraphics[width=0.86\textwidth]
{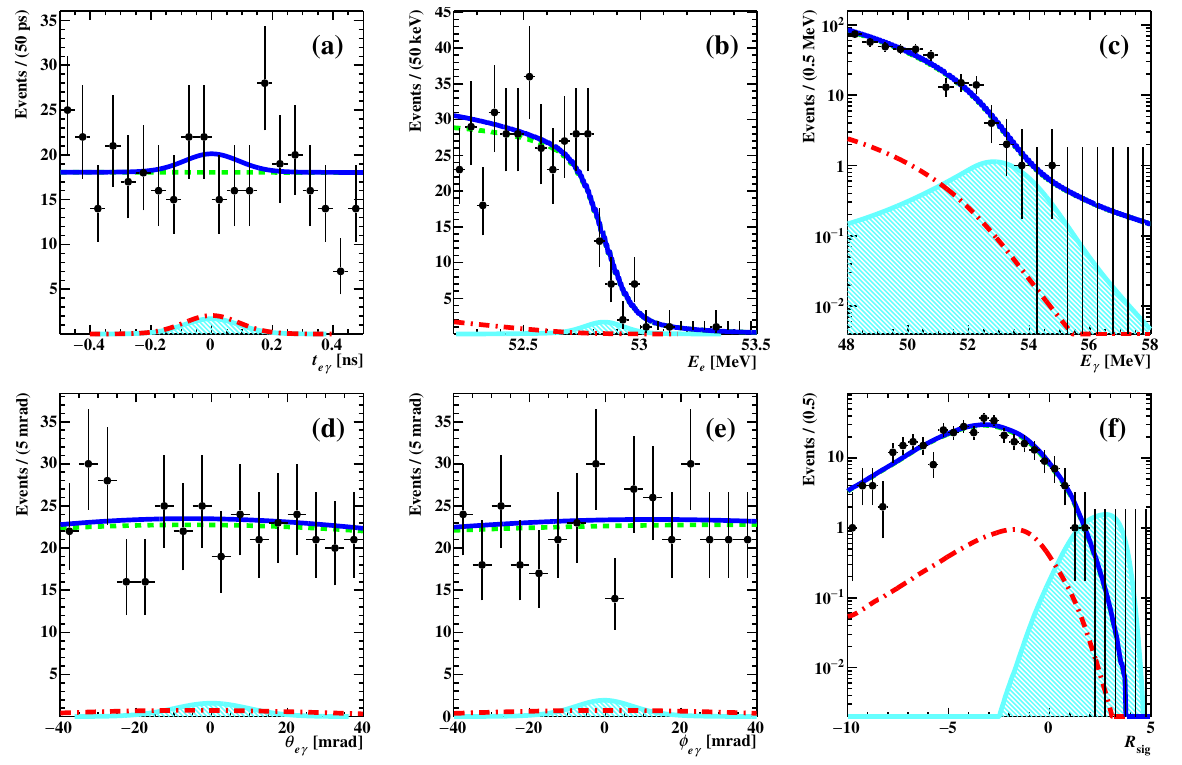}
\caption{
The projections of the best-fitted PDFs to the five main observables and $\rsig$, together with the data distributions (black dots). 
The green dash and red dot-dash lines are individual components of 
the fitted PDFs of ACC and RMD, respectively. The blue solid line is the sum 
of the best-fitted PDFs. The cyan hatched histograms show the signal PDFs corresponding to the four times magnified $\nsig$ upper limit.}
\label{fig:FitResult1D}
\end{figure*}

The MEG~II detector and its performance are described in \cite{MEGII:2023fog}. 
It consists of a positron spectrometer, formed by a cylindrical drift chamber (CDCH) and two  
semicylindrical sectors of each 256 scintillator pixels,
one located upstream of the target and the other downstream (pTC), 
placed inside a superconducting solenoid with a gradient magnetic field along the beam axis, and 
of a liquid xenon (LXe) photon detector, outside the solenoid, viewed by VUV-sensitive SiPMs on the front face 
and photomultiplier tubes (PMTs) on the other faces. 
In addition, a radiative decay counter (RDC) is located downstream
centred on the beam axis, to identify the ACC events with an
RMD-originated high-energy \photon-ray by tagging the corresponding 
low energy positron in coincidence. 
It consists of a scintillating plastic detector to measure the positron timing and a LYSO 
crystal calorimeter to measure the positron energy. 
The muon stopping target is elliptical,
its position $x_T$ and planarity are measured with a camera imaging dots printed on its surface and six holes bored in it.
A continuous monochromatic \muonp\ beam is delivered through a Beam Transport Solenoid onto the target where it is stopped.
The trigger for \megp\ events is based on the online estimates of the 
$\photon$-ray energy
$\egamma$, the relative time between the positron and the $\photon$-ray 
$t_{\positron\photon}$ from the LXe and the pTC, 
and the positron--$\photon$-ray relative direction from the same detectors.

A right-handed, Cartesian coordinate system is adopted, with the $z$-axis along the beam direction and the $y$-axis vertical and pointing upward. 
A polar spherical coordinate system is also used, with the $\theta$ angles referred as usual to the beam axis ($z$) and the $\phi$ angles lying in the $\left( x,y \right)$-plane.
For the LXe, we defined a local system of curvilinear coordinates $(u,v,w)$, where $u$ and $v$ are tangent to the cylindrical inner surface of the calorimeter (with $u$ parallel to $z$) and $w$ is the depth inside the LXe fiducial volume.
\section{Event reconstruction}

In each event, positron and \photon-ray candidates are described by five observables: $\epositron$, $\egamma$, $\tegamma$, $\thetaegamma$
and $\phiegamma$, $\thetaegamma$ and $\phiegamma$ being the 
polar and azimuthal angles of the relative direction, respectively.
The positron kinematics is reconstructed by tracking the trajectory in the magnetic spectrometer with the CDCH and extrapolating it backward to the decay vertex $(x_\positron, y_\positron, z_\positron)$
on the muon stopping target and forward to the pTC. 
The interaction time of the track with the pTC is measured with the hit times of close pixels combined in a 
cluster, to be corrected for the time of flight to obtain the positron production time $\tpositron$.

The tracking efficiency depends on the stopped muon rate on target
$R_\mu$, from $(66.0\pm 1.5 \pm 4.0_{R_\mu})\% $ at 
$R_\mu=$\SI{5e7}{\per\second} to $(77.0\pm 1.5 \pm 
4.0_{R_\mu})\% $ at $R_\mu=$\SI{2e7}{\per\second}, 
limited by the track finding capability.
Here, the ${R_\mu}$-subscripted uncertainty arises from the \SI{5}{\percent} uncertainty on $R_\mu$ \cite{CDCHPerfor,MEGII:2023fog}.
In addition, the positron efficiency is subject to the pTC acceptance and efficiency of $\SI{91(2)}{\percent}$. 
In the statistical analysis (see \sref{sec:Normalisation}), these efficiencies are naturally absorbed along with their uncertainties.

The conversion point of the incident \photon-ray inside LXe ($\ugamma$, $\vgamma$, $\wgamma$) and the conversion time are reconstructed by combining signals detected by sensors near the incident position.
The \photon-ray direction $(\thetagamma,\phigamma)$ is then reconstructed by joining ($\ugamma$, $\vgamma$, $\wgamma$) with the reconstructed decay vertex.
The \photon-ray conversion time $t_{\photon,\rm{LXe}}$ is corrected for the time of flight to obtain the production time $\tgamma$.
The resolution on $\tegamma = \tpositron - \tgamma$ is dominated by the time resolution of the LXe detector ($\sigma_{t_{\photon,\rm{LXe}}} = \SI{65}{\pico\second}$).
$\egamma$ is determined by summing the number of photons in all photosensors and converting it into energy by means of calibration factors accounting for the energy scale of the detector and its non-uniformity.
The RDC measures the time $t_{\positron,\rm{RDC}}$ and energy loss $E_{\positron,\rm{RDC}}$ 
of a low-energy positron in coincidence with a high-energy $\photon$-ray in the LXe detector.
\tref{perf} summarizes the detector performances used to build the Probability Density functions (PDFs) at $R_\mu = \SI{3e7}{\per\second}$ the most used value of stopped muon rate. 
For each value of $R_\mu$, the performances
change slightly and different PDFs are built.

\begin{table}
\caption{ 
Resolutions expressed in core Gaussian $\sigma$ and efficiencies of the MEG~II experiment in 2022 (2021) measured at $R_\mu = \SI{3e7}{\per\second}$.}
\centering
\newcommand{\minu}{\hphantom{$-$}}
\newcommand{\cc}[1]{\multicolumn{1}{c}{#1}}
\begin{tabular}{@{}lll}
\hline
  {\bf Resolutions }  & \minu \\ 
\hline\noalign{\smallskip}
$\epositron$ (\unit{keV})  & \minu 89 \\
$\phie,\thetae$ (\unit{mrad})& \minu 3.8/6.2 \\ 
$\ypos,\zpos$ (\unit{mm})   & \minu 0.61/1.76 \\ 
$\egamma$(\%)  ($\wgamma\SI{<2}{\cm}$)/($\wgamma\SI{>2}{\cm}$)  & \minu {2.4(2.0)/1.9(1.8)} \\
$\ugamma, \vgamma, \wgamma$ (\unit{mm})
& \minu {2.5/2.5/5.0} \\
$\tegamma$ (\unit{ps}) & \minu 78 \\
\hline
{\bf  Efficiencies (\%)} & \\ 
\hline
$\varepsilon_{\photon}$     & \minu 63 \\
$\varepsilon_{\positron}$   & \minu 67 \\
$\varepsilon_\mathrm{TRG}$  & \minu 91(88)  \\
\hline
\end{tabular}
\label{perf}
\end{table}

In this analysis, the reconstruction for 2021 dataset was updated from the one in \cite{MEGSearch2023}.
In the \photon-ray reconstruction, the analysis of pileup \photon-rays was revised 
\cite{PhotonPileupUpdate2024} 
to increase the signal efficiency by \SI{2}{\percent}.
In addition, \photon-ray calibration was updated relying on the \SI{17.6}{MeV} 
\photon-ray source \cite{MEGII:2023fog}, reducing the uncertainty in the energy 
scale due to temporal variation by \SI{40}{\percent}.
The positron selection was updated to select higher quality tracks,
improving the resolution for \SI{15}{\percent} for most of the variables.
The table reports the $\phie$ 
resolution averaged on $\phie$ while in \cite{MEGSearch2023} the
value for $\phi_e=0$ was reported.

\begin{figure}[tbp]
\centering
  \includegraphics[width=17pc,angle=0] {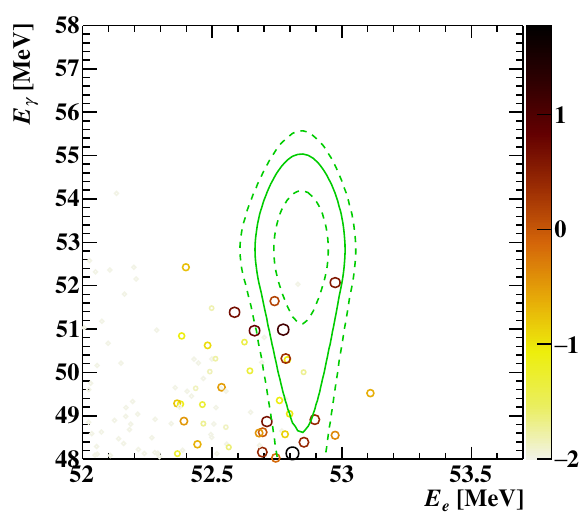}
  \includegraphics[width=17pc,angle=0] {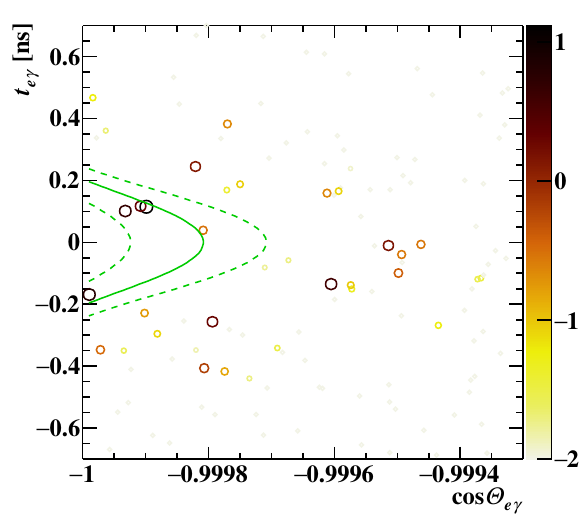}
 \caption{
Event distributions on the $(\epositron, \egamma)$- 
and $(\cos\Thetaegamma, \tegamma)$-planes with marker size and colour based on $\rsig$, where events with $\rsig<-2$ are clipped to be displayed as $\rsig=-2$. 
Selections of $\cos\Thetaegamma < -0.9995$ and 
$|\tegamma| < \SI{0.2}{\ns}$, which have \SI{97}{\percent} signal efficiency for each observable, 
are applied for the $(\epositron, \egamma)$-plane, 
while selections of $\num{49.0} < \egamma < \SI{55.0}{\MeV}$
and $52.5 < \epositron < \SI{53.2}{\MeV}$, 
which have signal efficiencies of \SI{93}{\percent} and \SI{97}{\percent}, respectively,
are applied for the $(\cos\Thetaegamma, \tegamma)$-plane.
The signal PDF contours ($1\sigma$, $1.64\sigma$ 
and $2\sigma$) are shown.
}
\label{fig:distribution2D}
\end{figure}

\section{Data taking}

The data analysed in this letter were collected in the years 
2022 (2021) for 18 (7) weeks,
for a total DAQ live time of \SI{7.8e6}{\second} (\SI{2.9e6}{\second}).
Data were taken at 
$R_\mu =$ \SIrange[range-phrase=\textendash,range-units=single,range-exponents=combine-bracket]{3e7}{5e7}{\per\second} 
(\SIrange[range-phrase=\textendash,range-units=single,range-exponents=combine-bracket]{2e7}{5e7}{\per\second})
for a total of \num{2.5e14}(\num{1.0e14}) stopped \muonp\ in 2022 (2021). 
The average fraction of the DAQ live time with respect to the total 
time was \SI{72}{\percent} (\SI{63}{\percent}) in 2022 (2021). 
During 2022, the fraction improved from 
\SI{60}{\percent} in the first 4 weeks to
\SI{80}{\percent} in the last ones.
These improvements were achieved thanks to the optimisation of the calibration DAQ scheme, 
which accounted for \SI{20}{\percent} and \SI{10}{\percent} of DAQ time at the beginning and end of 2022, respectively.

The DAQ efficiency, defined as the fraction of recorded events out of all events meeting the trigger requirements, was on average \SI{96}{\percent} (\SI{82}{\percent}) in 2022 (2021). 
This improvement was achieved thanks to the improvement of data recording bandwidth. 
The trigger logic was also improved thanks to the use of
PMTs instead of SiPMs in the online \photon-ray time 
reconstruction because the formers have faster time response.
As a result, the trigger efficiency for signal events improved from \SI{88}{\percent} in 2021 to \SI{91}{\percent} in 2022.

A degradation of the photon detection efficiency (PDE) of SiPMs in the LXe detector challenged the long-term operation \cite{MEGII:2023fog}.
Despite this, the average PDE was kept above \SI{10.8}{\percent} during
2022, thanks to annealing of SiPMs before the start of the data taking 
achieving an initial average PDE recovery to \SI{14}{\percent}.
On the other hand, the energy resolution in 2022 data worsened compared to 2021.
The cause of this degradation is under study, a possibility is lower calibration quality, mainly due to the non-linear response of SiPMs with the largest PDEs.
\begin{figure}[htb]
\centering
  \includegraphics[width=20pc,angle=0] {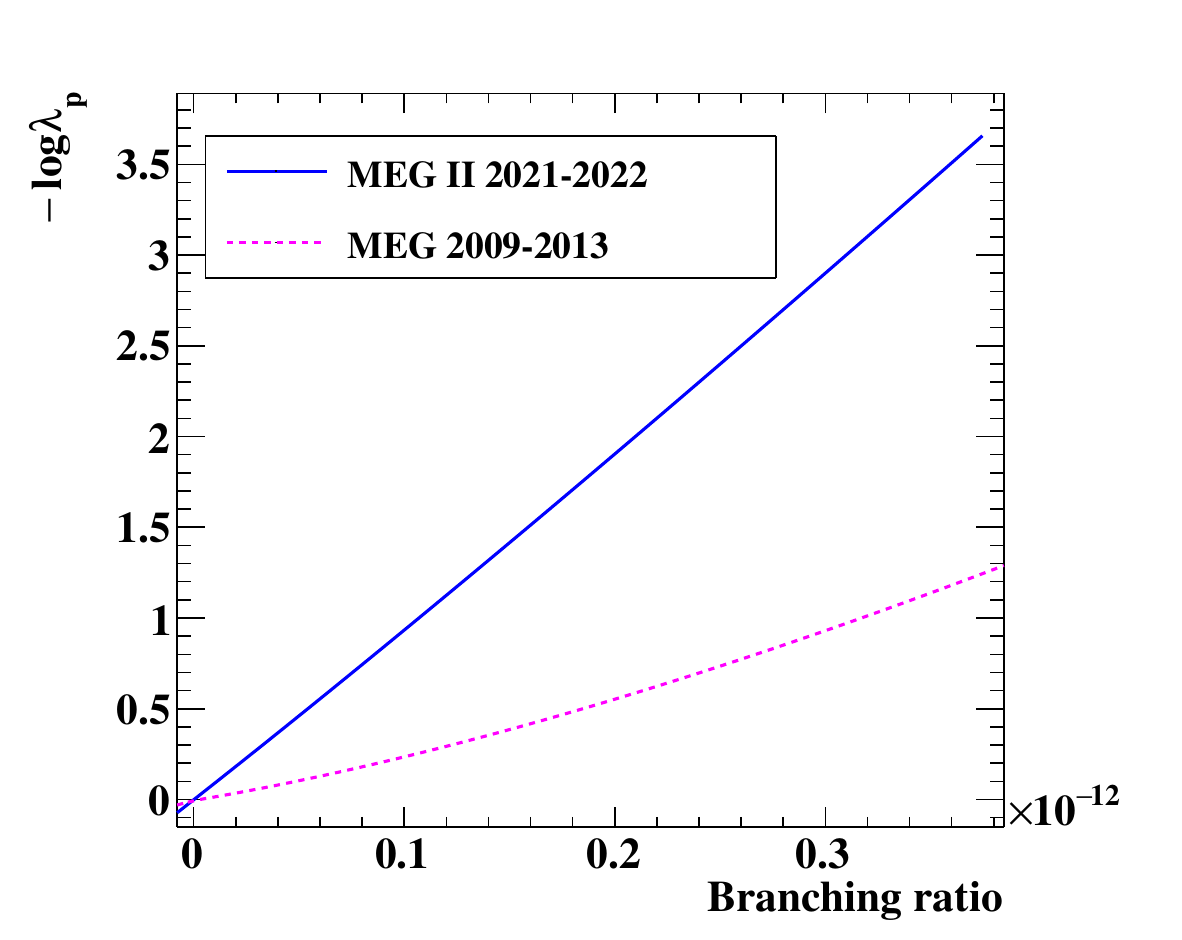}
 \caption{
 The negative log likelihood-ratio ($\lambda_{\rm p}$) versus the branching ratio. 
    The blue solid (magenta dashed) curve corresponds to the MEG~II 2021--2022 data (the MEG full dataset \cite{baldini_2016}). 
}
 \label{fig:NLL}
\end{figure}

\begin{figure}[htb]
\centering
  \includegraphics[width=20pc,angle=0] 
  {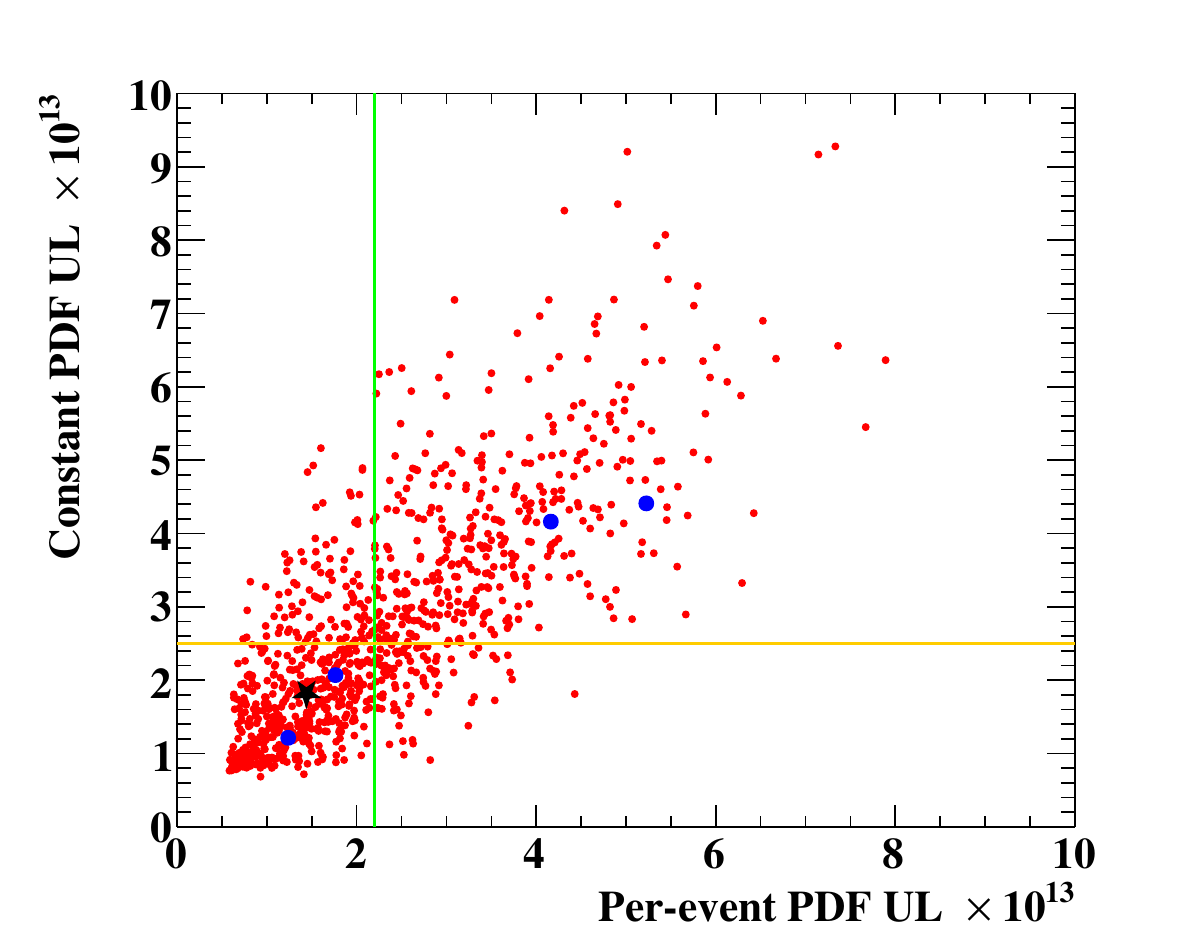}
 \caption{
Scatter plot of the 90 \%\ C.L. upper limits computed for
an ensemble of pseudo-experiments with a 
null-signal hypothesis for the two analyses. The blue dots and the black star are the upper limits measured in the time side-bands as described in the text and in the signal region, respectively. The vertical green and horizontal orange lines represent the sensitivities of the Per-event and Constant PDF analyses.  
}
\label{fig:sensitivity}
\end{figure}

\section{Analysis}
An extended unbinned maximum likelihood fit is performed to the dataset in the analysis region $\SI{48.0}{\MeV} < \egamma < \SI{58.0}{\MeV}$, $\SI{52.2}{\MeV} < \epositron < \SI{53.5}{\MeV}$, $|\tegamma|< \SI{0.5}{\nano\second}$, $|\phiegamma|< \SI{40}{\milli\radian}$, $|\thetaegamma| < \SI{40}{\milli\radian}$. 
The confidence interval for the number of signal events $\nsig$ is dictated by the Feldman--Cousins prescription~\cite{feldman_1998}, with ordering of the profile likelihood ratio \cite{PDG2024}. 
To translate the result into the branching ratio, 
a normalisation factor $N_\mu$, the number of effectively 
measured muon decays in the experiment,  is required $\BR(\megp) = \nsig/N_\mu$.
A blind analysis was performed, hiding events 
in the region $48.0<\egamma<\SI{58.0}{\MeV}$ and $|\tegamma|<\SI{1}{\ns}$,
until the analysis strategy was finalised.

In order to estimate the background the analysis was also applied to four fictitious analysis regions defining the time side-bands 
(\SI{-3.0}{\nano\second} $\tegamma< $ \SI{-2.0}{\nano\second}, 
\SI{-2.0}{\nano\second} $< \tegamma < $ \SI{-1.0}{\nano\second}, 
\SI{1.0}{\nano\second} $< \tegamma <$ \SI{2.0}{\nano\second}, 
\SI{2.0}{\nano\second} $< \tegamma< $ \SI{3.0}{\nano\second}.
Furthermore the region defined by $\SI{45}{\MeV} < \egamma < \SI{48}{\MeV}$  is called ‘‘$\egamma$ side-band used for the estimation of the RMD events.

A concise description of the analysis algorithm, which implements the approach  detailed in \cite{MEGSearch2023}, is presented in the following.

\subsection{Likelihood function}

A fit is performed with the set of observables
$\vec{x} = (\epositron, \egamma, \tegamma, \thetaegamma, \phiegamma, 
t_{\positron,\rm{RDC}} - t_{\gamma,\mathrm{LXe}},
E_{\positron,\mathrm{RDC}}, n_\mathrm{pTC})$.
The number of hits on the pTC, $n_\mathrm{pTC}$, is also introduced to incorporate the difference in its distribution between signal and background, as well as the dependence of $\tegamma$ resolution on $n_\mathrm{pTC}$.

The likelihood fit has four parameters: a floated parameter, $\nsig$, and three constrained parameters. $\nbg, \nrd, \xt$.
The likelihood function is obtained from PDFs for the signal ($S(\vec{x})$), RMD ($R(\vec{x})$) and ACC background events ($A(\vec{x})$) as
\begin{subequations}
   \begin{align*}
      &\mathcal{L}(\nsig, \nrd, \nacc, \xt) \nonumber =\\ 
      & \frac{e^{-(\nsig +\nrd+\nacc)}}{N_\mathrm{obs}!} 
      C(\nrd, \nacc, \xt) \times  \nonumber \\
      &\prod_{i=1}^{N_\mathrm{obs}} \bigl(\nsig S(\vec{x_i})+\nrd R(\vec{x_i})+\nacc A(\vec{x_i})\bigr) \; ,
      \label{eq:ExtendedLikelihood}
   \end{align*}
\end{subequations}

where $\vec{x_i}$ is the set of the observables for the $i$-th event;
$\nsig$, $\nrd$ and $\nacc$ are the
numbers of signal, RMD and ACC background events in the analysis region; 
$\xt$ is a parameter representing the misalignment of the muon stopping target;
$N_\mathrm{obs}$ is the total number of events observed in the analysis region;
$C(\nrd, \nacc, \xt)$ is a constraint term on the nuisance parameters, a product 
of three independent Gaussian functions, one for each parameter.
$\nrd$ is constrained by extrapolating the number of RMD events in the energy 
side-band, $\nbg$ is constrained by counting the number of accidental events in 
the timing sideband, and $\xt$ is constrained according to the 
estimated alignment precision.


The signal PDFs are modelled according to the measured resolutions including non-Gaussian tails.
The ACC $\epositron$ PDF is the convolution of the theoretical Michel spectrum with acceptance and resolution effects, fitted to data in the side-bands~\cite{CDCHPerfor}. 
The ACC $\egamma$ PDF is obtained from the Monte Carlo spectrum, 
with a Gaussian smearing and verified in the data.
The ACC angular PDFs are modelled with polynomials, fitted to data in the side-bands. 
The RMD PDFs are obtained by convolving the theoretical spectra with the experimental resolutions. 
The $n_\mathrm{pTC}$ PDFs are taken from the side-bands for the ACC background, and from the Monte Carlo for signal and RMD. 

\indent As a cross-check, two independent analyses are performed with differently constructed PDFs:
``per-event PDFs'' and ``constant PDFs''.
In the ``per-event PDF'' approach, the PDF parameters depend on the single event allowing the maximisation of the sensitivity.
Here, the event by event difference is modelled with parameters that represent the reconstruction performance for each event.
The positron PDFs are modelled with the tracking precision estimated by a Kalman filter technique.
The \photon-ray PDFs depend on the conversion position; 
the resolution of the \photon-ray measurement becomes worse when \photon-rays are converted near the end of the detectors fiducial volume.
In the other approach, ``constant PDFs'' are constructed by averaging such dependencies (except that different \photon-ray PDFs are used for $w_\photon < \SI{2}{\centi\meter}$ and $w_\photon > \SI{2}{\centi\meter}$).
Moreover, the angle PDF is also modelled differently:
the stereo angle $\Thetaegamma$ between the positron and the \photon-ray directions is used instead of the two separate components, $\phiegamma$ and $\thetaegamma$ and
the RDC observables are not used.
%
%
\subsection{Normalisation}
\label{sec:Normalisation}
The normalisation factor, $N_\mu = (1.34 \pm 0.07) \times 10^{13}$, 
is evaluated from the number of Michel positrons counted with a dedicated trigger~\cite{baldini_2016}.
The normalisation dataset is collected in parallel with the physics data-taking, such as to account for possible 
variations of the detector condition and of $R_\mu$.
For the purpose of this analysis, the 2021 dataset accounts for $N_\mu^{2021} = (0.28 \pm 0.01) \times 10^{13}$ with the trigger efficiency of \SI{88}{\percent}, which increased from \SI{80}{\percent} in \cite{MEGSearch2023} as a result of improvements in its evaluation. 
Among the factors contributing to the improvement, the most relevant are a more refined estimate of trigger direction match efficiency, driven by a better modelling of \photon-ray and positron behavior at the trigger level
, and an additional analysis selection that rejects events with a large expected energy deposit on dead channels.

\subsection{Results}

To evaluate the sensitivity, pseudo-experiments with a null-signal hypothesis were generated according to the PDFs and the evaluated number of background events from the side-bands.
With the ``per-event PDFs'' the median of the 
simulated \SI{90}{\percent} C.L. upper limit distribution
is $\sens = \num{2.2e-13}$ including systematic uncertainties. 
The contribution to the sensitivity due to systematic uncertainties, mainly due to detector misalignment, uncertainty on
detector position, $\photon$-ray energy scale and normalisation, is \SI{3}{\percent}.

A total of 357 events were observed in the analysis region.
The event distributions in each of $\epositron, \egamma, \tegamma, \thetaegamma, \phiegamma$, and $\rsig$ are shown in \fref{fig:FitResult1D}, 
where $\rsig$ is defined on the basis of \cite{PearsonNeyman1933} as 
\begin{equation*}
\rsig = \log_{10} \left( \frac{S(\vector{x}_i)}{f_\mathrm{RMD}R(\vector{x}_i)+f_\mathrm{ACC}A(\vector{x}_i)} \right) \; ,
\end{equation*}
with $f_\mathrm{RMD}$ and $f_\mathrm{ACC}$ being the fractions of the RMD and ACC background events, 
evaluated to be \num{0.027} and \num{0.973} in the side-bands, respectively. 
Scatter plots in the $(\epositron, \egamma)$ and $(\cos\Thetaegamma, \tegamma)$ planes are shown in~\fref{fig:distribution2D},
with the marker colour and size depending on $\rsig$.
The contours of the averaged signal PDFs are also shown.
No excess of events is observed in the signal region.
\fref{fig:NLL} shows the observed profile likelihood ratios as a function of the branching ratio for the 2021--2022 dataset of MEG II, compared with the full dataset of MEG. 
The best estimate of and the \SI{90}{\percent} C.L. upper limit on the branching ratio for the 2021--2022 MEG II dataset
are $\bestfit = \num{-3.8e-13}$ and $\ul = \num{1.5e-13}$, respectively.
This upper limit, that includes the systematic uncertainties, is consistent with the sensitivity 
calculated from the pseudo-experiments with a null-signal hypothesis.
Since the combination of MEG/MEGII data analysis improves the upper limit by a few \%\ only, we decided to not report a separate result.

The sensitivity of the analysis using the ``constant PDFs'' approach, evaluated with the same estimator, is $\sens = \num{2.5e-13}$ with 339 events in the analysis region.
The best estimate of and the \SI{90}{\percent} C.L. upper limit on the branching ratio, including systematic uncertainties,
are $\bestfit = \num{-5.0e-13}$ and 
$\ul = \num{1.9e-13}$.
This result is consistent with the expected \SI{\sim 15}{\percent} better sensitivity
of the ``per-event PDFs'' analysis.
Both analyses were also applied to four fictitious analysis regions inside the time side-bands 
($-3<\tegamma<\SI{-2}{\ns}, -2<\tegamma<\SI{-1}{\ns}, 1<\tegamma<\SI{2}{\ns}, 2<\tegamma<\SI{3}{\ns}$) 
and the results are consistent with the null hypothesis.
In \fref{fig:sensitivity} those results are shown together with the upper limits from an ensemble of pseudo-experiments; the vertical green and horizontal orange lines represent the sensitivities of the two analyses.

Finally, the likelihood fit in the analysis region was also performed without the constraints on 
$\nrd$ and $\nacc$. The best estimates of $\nrd=\num{0\pm 8}$ and $\nacc=\num{357 \pm 19}$ are consistent 
with the side-band estimates of $\nrd=\num{10.1\pm 1.7}$ and $\nacc=\num{364\pm 9.5}$, respectively.

\section{Conclusions and perspectives}

This letter presents the result of a blind, 
maximum-likelihood analysis 
applied to the data collected in 2021--2022 by the MEG II experiment with a sensitivity of $\sens = \num{2.2e-13}$; 
the improvement in sensitivity compared to \cite{baldini_2016} amounts to a factor 2.4.
The result is compatible with 
the expected background and establishes a \SI{90}{\percent} C.L. upper limit on the branching ratio
${\cal B} (\megp) < \num{1.5e-13}$, which is the most stringent to date.
The MEG~II collaboration has continued to take data during 2023 and 2024, with a
projected statistic 1.5-fold (1.3 (2023) + 0.2 (2024)) larger than in 2021--2022.
The collaboration plans to take data in the years 2025--2026 with an additional expected 
2.6-fold increase in statistics, with the goal of reaching a sensitivity to the $\megp$ decay of $\sens \num{\sim 6e-14}$.

\section*{Acknowledgments}

We are grateful for the support and cooperation provided  by PSI as the host laboratory and to the technical and engineering staff of our institutes.
This work is supported by 
DOE DEFG02-91ER40679 (USA); 
INFN (Italy); 
H2020 Marie Skłodowska-Curie ITN Grant Agreement 858199;
JSPS KAKENHI numbers JP26000004, 20H00154,21H04991,21H00065,22K21350 and JSPS Core-to-Core Program, A. Advanced Research Networks JPJSCCA20180004 (Japan);
Schweizerischer Nationalfonds (SNF) Grants 206021\_177038,
206021\_157742, 200020\_172706, 200020\_162654 and 200021\_137738 (Switzerland); the Leverhulme Trust, LIP-2021-01 (UK).
\bibliographystyle{my}
\bibliography{MEG}

\end{document}